# Evaluation of Single-Chip, Real-Time Tomographic Data Processing on FPGA - SoC Devices

G. Korcyl*, P. Białas, C. Curceanu, E. Czerwiński, K. Dulski, B. Flak, A. Gajos, B. Głowacz, M. Gorgol, B. C. Hiesmayr, B. Jasińska, K. Kacprzak, M. Kajetanowicz, D. Kisielewska, P. Kowalski, T. Kozik, N. Krawczyk, W. Krzemień, E. Kubicz, M. Mohammed, Sz. Niedźwiecki, M. Pawlik-Niedźwiecka, M. Pałka, L. Raczyński, P. Rajda, Z. Rudy, P. Salabura, N. G. Sharma, S. Sharma, R. Y. Shopa, M. Skurzok, M. Silarski, P. Strzempek, A. Wieczorek, W. Wiślicki, R. Zaleski, B. Zgardzińska, M. Zieliński, P. Moskal

*Abstract*—A novel approach to tomographic data processing has been developed and evaluated using the Jagiellonian PET (J-PET) scanner as an example. We propose a system in which there is no need for powerful, local to the scanner processing facility, capable to reconstruct images on the fly. Instead we introduce a Field Programmable Gate Array (FPGA) System-on-Chip (SoC) platform connected directly to data streams coming from the scanner, which can perform event building, filtering, coincidence search and Region-Of-Response (ROR) reconstruction by the programmable logic and visualization by the integrated processors. The platform significantly reduces data volume converting raw data to a list-mode representation, while generating visualization on the fly.

## I. INTRODUCTION

TOMOGRAPHIC image reconstruction algorithms were introduced in 1960's [1]. Many sophisticated methods have been developed since, focusing mainly on delivering high quality images, as fast as possible. Those methods involve heavy computational iterative procedures like Maximum Likelihood-Expectation Maximization (MLEM) [2] and accelerated variations e.g. Ordered Subsets MLEM (OSEM) [3]. Data processing systems have been developed accordingly, providing more and more computing power in order to meet the growing data volumes and algorithms complexity [4].

Modern trends in nuclear medical imaging introduce whole-body scanners with three-dimensional (3D) acquisition mode, where the Field-Of-View (FOV) is extended from typical 20 cm to almost 200 cm [5,6]. Such extension must result in a proportional increase of the generated data volume and therefore required processing power.

G. Korcyl* and P. Białas are with the Department of Information Technologies at Faculty of Physics, Astronomy and Applied Computer Science, Jagiellonian University, Kraków, Poland (e-mails: grzegorz.korcyl@uj.edu.pl, pbialas@th.if.uj.edu.pl)

E. Czerwiński, K. Dulski, A. Gajos, B. Głowacz, K. Kacprzak, D. Kisielewska, T. Kozik, N. Krawczyk, E. Kubicz, Sz. Niedźwiecki, M. Pawlik-Niedźwiecka, M. Pałka, Z. Rudy, P. Salabura, N. G. Sharma, S. Sharma, M. Skurzok, M. Silarski, M. Zieliński and P. Moskal are with the Faculty of Physics, Astronomy and Applied Computer Science, Jagiellonian University, Kraków, Poland (e-mails: eryk.czerwinski@uj.edu.pl, kamil.dulski@gmail.com, alek.gajos@gmail.com, bartoszglowacz@gmail.com, k.kacprzak@uj.edu.pl, dk.dariakaminka@gmail.com, ufkozik@cyf-kr.edu.pl, nikodem.krawczyk@gmail.com, ewelinakubicz1@gmail.com, szymonniedzwiecki@gmail.com, monikapawlik88@gmail.com, marek.palka@gmail.com, ufrudy@cyf-kr.edu.pl, piotr.salabura@uj.edu.pl, neha.gupta@uj.edu.pl, sushil.sharma.uj@googlemail.com, magdalena.skurzok@uj.edu.pl, michal.silarski@uj.edu.pl, marcin.zielinski@uj.edu.pl, ufmoskal@if.uj.edu.pl).

C. Curceanu is with INFN, Laboratori Nazionali di Frascati, Frascati, Italy (e-mail: catalina.curceanu@lnf.infn.it).

B. Flak and P. Rajda are with the Faculty of Computer Science, Electronics and Telecommunications, AGH University of Science and Technology, Kraków, Poland (e-mails: flakbartlomiej@gmail.com, pjrajda@agh.edu.pl)

B. C. Hiesmayr is with Faculty of Physics, University of Vienna, Vienna, Austria (e-mail: beatrix.hiesmayr@univie.ac.at).

M. Gorgol, B. Jasińska, R. Zaleski and B. Zgardzińska are with the Institute of Physics, Maria Curie-Skłodowska University, Lublin, Poland (e-mails: marek.gorgol@gmail.com, bozena.jasinska@poczta.umcs.lublin.pl, radek@zaleski.umcs.pl, bozena.zgardzinska@poczta.umcs.lublin.pl).

M. Kajetanowicz is with Nowoczesna Elektronika Sp. Z o.o. Kraków, Poland (e-mail: mkajetan@ne.com.pl).

P. Kowalski, W. Krzemień, L. Raczyński, R. Y. Shopa and W. Wiślicki are with the Department of Complex Systems, National Centre for Nuclear Research, Otwock-Świerk, Poland (e-mails: pawel.kowalski@ncbj.gov.pl, wojciech.krzemien@ncbj.gov.pl, lech.raczynski@ncbj.gov.pl, roman.shopa@ncbj.gov.pl, wojciech.wislicki@ncbj.gov.pl).

M. Mohammed is with the Faculty of Physics, Astronomy and Applied Computer Science, Jagiellonian University, Kraków, Poland and the Department of Physics, College of Education for Pure Sciences, University of Mosul, Mosul, Iraq (e-mail: mohsen_albadrani@yahoo.com).

P. Strzempek was with the Faculty of Physics, Astronomy and Applied Computer Science, Jagiellonian University, Kraków, Poland (e-mail: pawelstrzempek@interia.pl).

A. Wieczorek is with the Faculty of Physics, Astronomy and Applied Computer Science, Jagiellonian University, Kraków, Poland (e-mail: anna.m.wieczorek215@gmail.com).

In this paper, we present a solution for this challenge. Instead of expanding the processing system of the scanner, the goal is to replace it with a compact and integrated, FPGA based module. High-end System-On-Chip (SoC) devices, which are FPGA, CPU and GPU enclosed in a single chip [7] provide enough computational power in order to produce preliminary image in real-time. Programmable logic is perfectly suitable for the implementation of Event-By-Event, incremental reconstruction algorithms. Preprocessed data, in a form of points cloud or list-mode data can be delivered to the services located in the cloud for a full-featured, high-quality reconstruction, still within reasonable amount of time. In this way we save space, weight and costs required by complex computing platforms.

*A. Background*

Reconstruction algorithms for the whole-body and 3D PET imaging impose tremendous requirement on memory in order to store the system matrix and voxelated volume [8]. Therefore there is a demand for exploring alternative 3D image representations such as point clouds and tetrahedral mesh [9], which are more suitable for three-dimensional structures.

Many successful projects [4,10] employ CPUs and GPUs for algorithms that require a particular portion of the data to be recorded and then start the reconstructing procedure. Such systems do not operate in the real-time regime, even though an impression of "live" visualization of the measurement can be achieved. The data from the scanner has to be stored in memory and then requested by the operating system. This deviates from the real-time processing path as the time between data reception and analysis becomes non-deterministic.

During last few years, the FPGA technology has considerably advanced offering devices with very high amount of resources (quadrupled since 2012) [7] and firmware development methodologies accelerating algorithms implementation. Device families, optimized for particular applications can be used at various stages of the readout system. Xilinx Virtex Ultrascale+ family provides unparalleled performance when it comes to process multiple (up to 128) data streams. Whereas Xilinx Zynq MPSoC (MultiProcessor System-on-Chip) family is a hybrid housing FPGA resources, a quad-core ARM A53, dual-core ARM R5 and a Mali-400 GPU inside a single chip. Those devices are perfect for the implementation of high level algorithms and visualization. It is worth to mention that ARM processors are more often considered when it comes to High-Performance Computing (HPC) [11] as they provide reasonable computing power at ultra-low power consumption. Therefore it is justified to anticipate in the near future devices capable to preprocess data in programmable logic and perform full image reconstruction by the integrated processors.

Scanner modules can be composed of light-weight, plastic scintillators with silicon photomultipliers. Combined with ultra-low-power and integrated instant image reconstruction can yield a new class of scanning devices that are modular and portable. Such systems can find application in Image-Guided Surgery (IGS) as devices supporting clinical procedures [12] and could enhance dosimetry in treatment with proton beam providing feedback to control the beam properties [13]. Both applications require maximum responsiveness of the visualization mechanism. Real-time access to reconstructed data provides a way to improve motion correction procedures [14] or monitoring of physiological processes dynamics.

*B. Overview of Tomographic Data Processing*

Electronic readout systems digitize the signals generated by the detecting material of the scanner. Precise timestamp of the signal and its charge are required for image reconstruction. For this purpose Analog-to-Digital Converters (ADC) are most commonly used. They sample the signal in the analog domain at fixed frequency, delivering a series of values, from which one can reconstruct the analog signal shape. Another way of measuring those values is to employ Time-to-Digital Converters (TDC) and Time-Over-Threshold (TOT) method [15]. TDC discriminates an analog signal at some amplitude threshold level and precisely registers the time of the leading edge (starting time) and width of the signal.

A set of data from the entire scanner has to be collected and time coincidences between registered hits have to be found in order to recover potential LORs (Fig. 1). Dedicated coincidence processors are often developed in order to retrieve such conditions in real-time [16]. There are also efforts to implement such search on the software side [4].

A result of coincidence search is a list of detector channels that registered a signal, referred to as list-mode data structure, which contains localization and timing information. Such lists are then processed by image reconstruction algorithms.

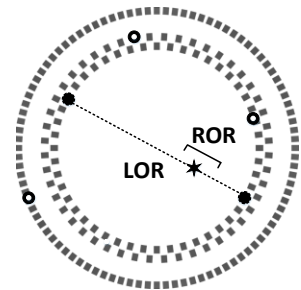

Fig. 1. Schematic, front view of J-PET scanner with 3 detector layers. ROR reconstruction process is composed of 3 steps: a) Annihilation photons are emitted from a particular place (black star) in opposite directions and hit the detector strips. B) The hits are registered (black points), together with possible noise e.g. due to the scattering in the detector (black open circles) that needs to be suppressed. Two hits on different detectors, within a defined time window are LOR candidates. C) Time difference between the hits on two strips (TOF) is used to determine a section along LOR from which the gamma quanta originated, this section in 3D space is a ROR.

State-of-the-art image reconstruction algorithms exploit Expectation Maximization [17] techniques which are statistical methods to estimate the radioactivity density distribution. They require multiple iterations over entire data sets in order to approximate the intensity map, until some quality condition is met.

Large data sets, especially when it comes to highly-granular and wide-FOV detecting systems create a significant performance problem for those methods [8]. Alternative methods like Origin Ensemble [18] attempts to tackle this issue but still rely on iterative approach for image reconstruction, therefore remain not suitable for incorporation into real-time processing path.

There is a number of researches that approached the problem of image reconstruction in real-time. Most of them are based on CPU-GPU computational platforms that process the data outside of the DAQ chain, therefore leaving the true real-time data path [19,20]. Other solutions engage FPGAs for high-level data processing [21,22]. However, so far a true real-time image reconstruction, running with a full scale scanner was not achieved. FPGA devices are often used as signal processing units, coincidence finders or off-line reconstruction accelerators [23]. The uniqueness of the solution described in this article is that it incorporates all functions needed for image reconstruction on a single chip, that process the data as it flows through the system, delivering instant image generation, without any loss of data.

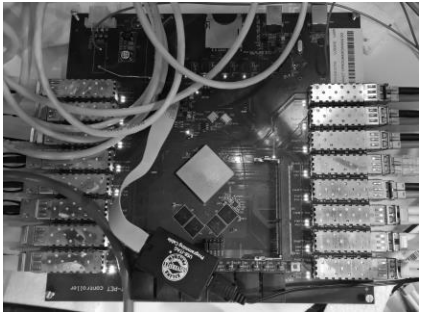

Fig. 2. Photo of the J-PET Controller board. The left set of 8 optical links are the inputs from the digitizing boards. The streams are being processed by the logic implemented on centrally located Zynq device. The right set of 8 optical links is used to transmit raw or list-mode data streams and algorithm results to the computer.

*C. Proposed Solution*

We have designed and constructed J-PET Controller (Fig. 2), a hardware platform for processing data from the scanner. The board consists of Xilinx Zynq device, 16 optical transceivers and DDR3 memory.

Processing firmware has been developed and evaluated with the use of Jagiellonian PET (J-PET) scanner prototype [24-29] (Fig. 3), which is the first 3D TOF Positron Emission Tomography scanner built of plastic scintillators having axially arranged strips forming cylindrical diagnostic chamber. The tests show that the processing platform can process multiple data streams, extract Lines-of-Response (LORs), calculate Regions-of-Response (a section of LOR using TOF information [30]) and generate a basic visualization of the collected data.

This work is an early proof-of-concept and defines a clear development roadmap towards single-chip, integrated and compact processing solution.

In the following sections, the system under discussion is presented in detail. In section II, we describe how the data is produced, what is its content and how the DAQ system works. Section III explains the processing algorithm and steps required to perform in order to calculate RORs from the input data streams. Implementation details are enclosed in section IV. Laboratory tests for evaluation of the developed solutions in real environment were performed and are described in section V. The paper concludes with an overview of possible improvements that can be implemented and a summary in the last section VI.

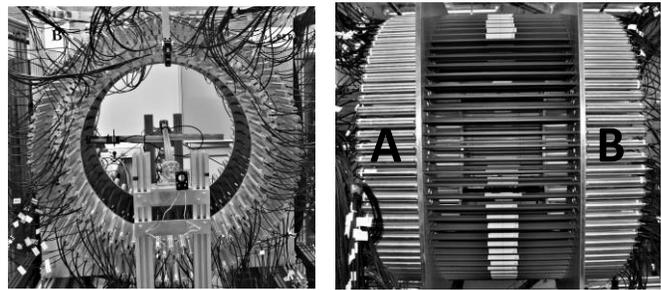

Fig. 3. J-PET scanner photos. Left: front view of the scanner. 3 layers of detector strips are visible with cabling to high voltage and data acquisition system. A rotating arm is placed in the center of the scanner and is used for various tests and calibration. Right: a side view of the scanner. Detecting modules consist of a plastic scintillating strips and two photomultipliers attached: one on the left side (A) and the other on the right side (B). The inner diameter of the diagnostic chamber is equal to 85 cm and the length of the scintillators is 50 cm [29].

## II. DATA ACQUISITION SYSTEM AND DATA STRUCTURE

In order to fully understand the nature of the data that is being processed by the system, a detailed J-PET detector structure, its data acquisition system and the implemented readout procedure are presented briefly below (for full review see [31,32]).

The detector constructed at the Jagiellonian University is composed of 192 modules, built from plastic scintillators, arranged into a barrel with 3 layers (see Fig. 1 and Fig. 3) [28]. The inner and the middle layer consist of 48 modules each and the outer one of 96 modules. Each module is 0.5 m long and has two photomultipliers (PMT) attached to its ends. This gives a total 384 analog signal sources to process.

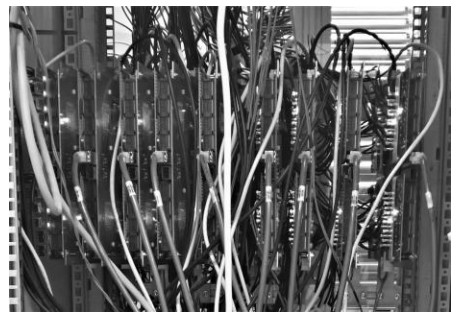

Fig. 4. Photo of the TRB system designed for the J-PET scanner. A custom create houses 9 TRB boards interconnected with optical links. 384 analog cables are connected from the back side.

*A. Architecture of DAQ System and Readout Procedure*

All those signals are delivered to the digitizing system based on the Trigger Readout Board (TRB) platform [33,34] (Fig. 3 and Fig. 4). It contains 4 peripheral FPGAs, hosting 48-channel Time-to-Digital Converter (TDC) each and one central FPGA for the TDC readout and data transmission. The modules use TDCs to measure the time of arrival and the width (which allows to estimate the collected charge of the analog signal) of signals generated by the Leading Edge Discriminator (LED) with a high time resolution of 12 ps [35]. The LEDs are placed between the PMTs and the TDCs. Each analog signal is sampled in the voltage domain at four thresholds by the dedicated FPGA based Multi-Voltage Threshold front-end [36]. This gives us 4 points on the leading edge and 4 points on the trailing edge of the analog signal, what allows to reconstruct the original signal shape [37-39]. Consequently, to process the signals for the whole detector 4 (thresholds) x 2 (sides) x 192 (strips) = 1536 TDC channels, therefore 8 TRB boards are required. One additional master module controls the readout procedure, synchronizes all the slaves and acts as a gateway for control and monitoring messages exchange. The architecture of this system is shown in Fig. 5.

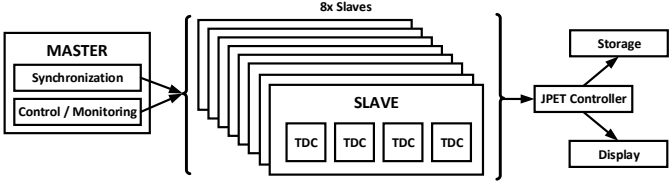

Fig. 5. Schematic view of the J-PET readout system. One master module communicates with slaves and synchronizes the readout procedure. Each slave registers the hit times, measured with the TDC devices and transmits collected data via dedicated GbE connection to the J-PET Controller, which performs the processing delivering either raw or list-mode data to the storage and a visualization of analyzed data.

Each slave module collects the data from its TDCs and sends it out using 1 Gigabit Ethernet (GbE) network for further processing to J-PET Controller (described in more details in section V). Looking at the system from the data processing perspective, we have 8 data stream sources that have to be analyzed.

An important aspect is the readout procedure which defines the data stream characteristics and its content. In the J-PET scanner, we have applied a continuous readout scheme [32]. That means that the system constantly measures, collects and transmits data. This is in contrary to a triggered system, which reacts only upon meeting some predefined conditions. Continuous readout allows to collect more detailed data but at a cost of much higher data volume, due to high overhead (even when no hits were registered, data packets with headers only are generated) and lack of preliminary selection. The implementation of such readout procedure in our system is realized by the master module, which sends readout request messages to all the slaves in a synchronized manner and with a constant rate of 50 kHz. Each slave measures the leading and trailing times of all generated signals with respect to the common start time, stores them in a buffer and sends them using Gigabit Ethernet network. The system splits the entire measurement period into 20 µs *timeslots* by synchronously initiating readout of all slaves, at 50 kHz rate. That means that in order to reconstruct the scanner state during one such timeslot, one has to collect and combine together 8 data packets containing the data from the same 20 µs time epoch, marked by the timeslot number. From that point on, it is possible to perform analysis on a higher level.

A similar concept has been applied in a software-based coincidence engine described in [40]. Although the concept of measurement fragmentation into fixed-length time intervals is the same, our system is designed to process them directly as they are received, with a deterministic latency. In software solution, the data units are stored in a processing queue and accessed by an available thread at an arbitrary time.

*B. Timeslot Content*

Single timeslot contains data that represents all registered signals from the entire detector during particular 20 µs period of the measurement. Such 50 kHz frequency has been chosen in order to record most of the data, taking into account hit rates on channels, buffering capabilities of the slave TRB modules and GbE gateways throughput [41]. As interesting events (hits on two detectors for possible LOR) happen in a very short time, in range up to few nanoseconds, we examine each timeslot independently from the others, considering that the number of events that span over two consecutive timeslots is negligible. Such approach gives us a great advantage, because we divide the entire measurement into equally long timeslots that are being delivered for processing at a constant frequency.

Registered signals are represented by 32-bit data words generated by the TDC for each channel: one for the leading edge and one for the trailing edge time. The measured time is a combination of three components: fine time measurement in the range from 0 to 5 ns with 12 ps binning, coarse time measurement in the range from 0 to 10.235 µs with 5 ns binning and epoch counter in range from 0 to 45.8 min with 10.235 µs binning. Those three values have to be combined in order to get an absolute time of a hit. The word containing epoch counter is inserted only when the coarse counter overflows and there is a hit to be registered. Fine and coarse times, together with the channel number are composed into one 32-bit word. The epoch counter occupies another 32-bit word.

## III. PROCESSING ALGORITHM

In order to reconstruct a tomographic image one has to process timeslots and accumulate enough statistics to display well pronounced regions of interest against background. The processing is divided into several steps:

1) *Reassembly of data units into timeslots*
2) *Extraction of hit times*
3) *Mapping of the detector geometry*
4) *Application of calibration parameters*
5) *Coincidence search*
6) *Calculation of ROR parameters*
7) *Histogramming of RORs*

System components performing the above steps are described in detail in following subsections.

*A. Decomposition Channel*

The continuous readout mode of the system, triggers the digitizing slave boards to transmit out the current timeslot exactly at the same time. Depending on the size of particular packet (time required for packet construction is linear to the payload size), slight offsets between data packets arrival on the receiving side appear. This effect is compensated by the use of derandomizing FIFO buffers right after payload reception.

The timeslots are being processed individually, one after the other. That gives a fixed time span of 20 µs during which the buffers have to be emptied. From now on, we will discuss the processing of such single timeslot as the entire image reconstruction is a statistic built on RORs extracted from multiple timeslots.

Before the data fragments get reassembled into a complete timeslot, we can process each fragment individually. There are 8 instances of the decomposition channel (Fig. 6), one per input data stream, all processing in parallel. Each channel includes an User Datagram Protocol (UDP) stack in order to receive data packets (GbE Receiver) and store the payload in the Derandomizing Buffer.

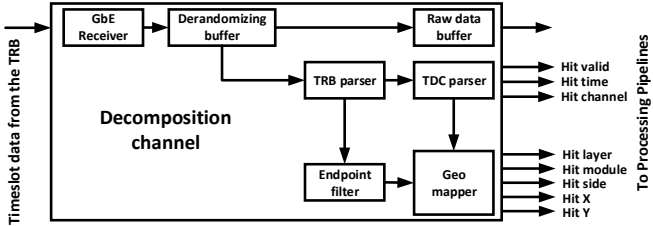

Fig. 6. Block diagram of the logic included in the decomposition channel. Raw data from the TRB board flows through the modules which perform hit extraction and geometry mapping. The raw data is kept in a separate buffer providing access to the original data packets at any time. Extracted information is processed further by Processing Pipelines.

When data is available in all Decomposition Channels, the payload is duplicated to the Raw Data Buffer and to the TRB Parser that extracts the timeslot number for synchronization, device ID for proper channel mapping and the registered data. The data words with hit times contain the TDC channel number, which together with the device ID is used by the Geometry Mapper to assign both: Layer (1-3), Side (Left– A, Right – B) and Strip (1-48 for Layers 1 and 2, 1-96 for Layer 3) as well as its X and Y coordinates. In the same time, three components (fine, coarse and epoch) of the hit time are calculated into a single, absolute time value by the TDC Parser. Such absolute values are then synchronized together using time markers that signal the start of a timeslot and are registered by the TDCs on dedicated reference channels. It is also a place where calibration parameters and applied for TDC effects such as Differential Non-Linearity (DNL) [42] and channel to channel time offsets. The times are adjusted to a timeslot range that is 0 to 20 μs. Additionally the leading and the trailing edge times of a single hit are combined together into values representing the time of arrival and the width. All those actions are performed in a streaming way, that means no additional buffering is needed and no deadtime is introduced.

### B. Processing Pipelines

Data from all decomposition channels is combined into one stream and delivered to processing pipelines (Fig. 7) which implement algorithms that are executed on hit data stream. All processing pipelines are instantiated in parallel with respect to each other and their processing stages are decomposed into several functional modules. The modules can be of two forms: non-buffered and buffered. The first one can only use registers that introduce couple of cycles latency. While the second modules store the incoming data stream in a FIFO buffer.

The modules can exchange information between themselves and between the pipelines. In this way, the entire processing flow can be synchronized and pipeline modules can use products calculated by other pipelines.

In current design two processing pipelines are implemented. One is the coincidence search engine that finds potential LOR candidates. The second calculates ROR points coordinates.

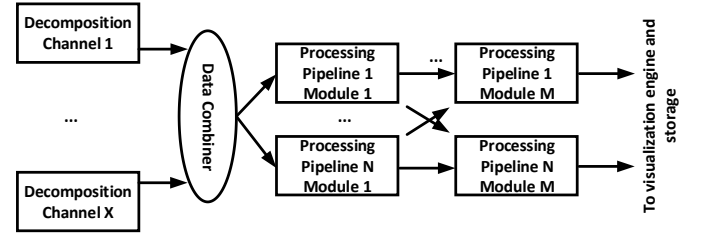

Fig. 7. Block diagram presenting the processing flow. Output from a number of decomposition channels is combined into a single data bus and then distributed to a number of processing pipelines (two in the current version of J-PET). Each processing pipeline performs an algorithm and can consist of many pipelined modules. The modules can communicate between themselves in order to synchronize the processing flow.

### C. Coincidence Search Pipeline

Hit times, together with the scintillator strips coordinates, are sufficient to compute the RORs. The only condition that can indicate a possible ROR is the fact that two detector strips have registered a hit in a relatively short time interval, less than 100 ns, which is the maximum diagonal time of flight of the gamma quanta to scintillator, including light propagation to PMT plus analog processing time and a margin. It means one pair of hits (H1 and H2) have to be present in the data (Fig. 8).

Each decomposition channel processes a particular segment of the J-PET (see Table I). Having that information, it is possible to continue parallel and streamlined processing. In order to parallelize processing even more, one timeslot is divided into 32 fragments, each 625 ns long, called *timebins* (Fig. 8).

TABLE I
DETECTOR SEGMENTATION

| Decomposition channel | J-PET segment | Single Strip Coincidence |
|---|---|---|
| 1 | Layer 1 x Side A x Strip 1 - 48 | AND |
| 2 | Layer 1 x Side B x Strip 1 - 48 | |
| 3 | Layer 2 x Side A x Strip 1 – 48 | AND |
| 4 | Layer 2 x Side B x Strip 1 - 48 | |
| 5 | Layer 3 x Side A x Strip 1 - 48 | AND |
| 6 | Layer 3 x Side B x Strip 1 – 48 | |
| 7 | Layer 3 x Side A x Strip 49 – 96 | AND |
| 8 | Layer 3 x Side B x Strip 49 - 96 | |

The search for coincidence is performed by setting bits active in 2D arrays timebins x Strips in case the time of arrival of a hit is registered by a particular Strip, within particular timebin. Such array is constructed for each decomposition channel. The hit data flows through the pipeline and the arrays are immediately updated with information about the time and location of the detector channels that fired.

When there is no more data in the derandomizing buffers it means that the content of the entire timeslot has been processed, the arrays are complete and coincidences can be found. First, single-strip (Side A and Side B on the same Layer and Strip) are calculated. Such operation is a logic AND of two binary arrays (Fig. 9), which on FPGA is performed within 1 clock cycle. It is realized for Layer 1, Layer 2 and Layer 3 in parallel (Table I). That means that after the data flows through the coincidence search module, it requires exactly one clock cycle to find all detector strips that have coincidence hits on both sides and are potential candidates for LORs.

Second clock cycle is required to find if there are two or more detector strips that fired within a particular timebin. This operation is performed at the same time for all timebins.

Third clock cycle is used to construct the output result that is information if the current timeslot contains LOR candidates, on which detector channels and within which timebin.

Additional 2 clock cycles are used to register the input and output data vectors for the coincidence search module. That means that 5 clock cycles, at 200 MHz clock it is 25 ns, are needed to search 8 arrays, each 32 (timebins) x 48 (channels) elements. On standard CPU, one would require 6 nested FOR statements in order to iterate over all elements in the similar manner and additional time needed for memory accesses. Recent computing platforms would consume 2 or 3 orders of magnitude more time to realize such task.

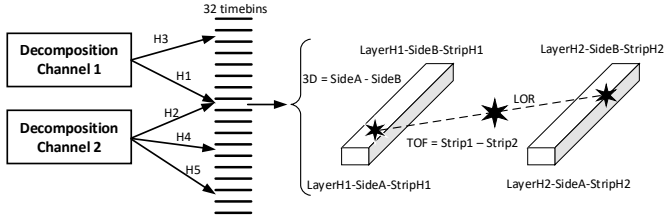

Fig. 8. Coincidence search pipeline data flow. Hits extracted by the decomposition channels are assigned to a particular timebin depending on their time within a timeslot. In the presented example, hits H1 and H2 are in time coincidence while hits H3, H4 and H5 are classified into distinct timebins and cannot form coincidence. All timebins are then processed in parallel and LOR candidates are determined. In order to get the $3^{rd}$ dimension coordinate, time difference between two sides of a single strip has to be calculated. TOF is a time difference between hits on two strips.

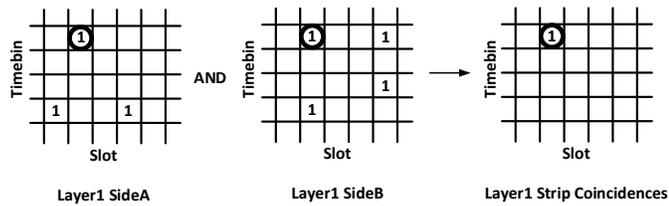

Fig. 9. Example arrays constructed on decomposition channel 1 (processing data from Layer 1 Side A) and decomposition channel 2 (data from Layer 1 Side B). While the hits are being processed, the arrays are filled with active bits, corresponding to hits on particular detector Slot within particular timebin. A logic AND operation leaves the array elements active (circled), only in case if there are hits in the same detector Slot within particular timebin, which gives all the single-strip coincidences.

### D. ROR Calculator Pipeline

As data flows from the decomposition channel to the coincidence search module it is duplicated and enters a second processing pipeline. Eight buffers store extracted hit information. Once the current timeslot data is entirely readout and coincidence search module signalizes that LOR candidates search is finished, the hits are being readout from the buffers and filtered by the application of timebin mask. Only hits within a timebin, which has been qualified by the coincidence search stream for containing potential LOR are being transferred further. All the other hits are being treated as noise and dropped.

Accepted hits, depending on their time of arrival, are directed to one of 32 ROR processors, one per timebin and stored in a RAM memory block. When the last one arrives, the ROR processor begins, in an iterative way, pairing hits each vs each. For each such pair an additional time difference filter of less than 10 ns is applied. For pairs that were positively qualified, annihilation coordinates are calculated using timing information to determine registration point along the strips (Z axis) and between the strips (TOF), as presented in Fig. 8. Set of 3 coordinates values: X, Y and Z, is stored in the output FIFO buffer.

### E. ROR Histogramming

Four ROR packagers iterate in a round-robin way through the ROR calculators output buffers and stream the coordinates to a shared DDR memory, that is accessible through the integrated ARM processor. A dedicated Linux distribution PetaLinux runs software that reads calculated points coordinates, builds a 2D histogram and 3D point cloud representation of acquired data and makes it available to access through a web server.

The same data set can be sent through output optical links to external storage in a form of list-mode coincidences, making it possible to reconstruct with offline software algorithms.

## IV. IMPLEMENTATION

The algorithm described above has been implemented on a single Xilinx Zynq XC7Z045-3FFG900. All its components (except visualization) are implemented in Programmable Logic (PL) resources. The design is highly configurable at synthesis time by generic parameters, through which one can set the number of decomposition channels, assign addresses and define the number of timebins which influences the most resource usage.

Core components are written as RTL in VHDL. The communication infrastructure, that is Gigabit Ethernet transceivers with UDP stack are ported from Lattice ECP3 implementation [33] and processed with native 8-bit wide data buses at 125 MHz. 32-bit bus with a 200 MHz clock is the output from the decomposition channel and is common for the rest of the design components.

Some components such as detector geometry mapper and ROR calculator were implemented using Vivado High Level Synthesis [43].

In Table II resource usage for the entire design, including 8 decomposition channels and 32 timebins is reported. Because of the heavy emphasis on true real-time processing most critical resource are Look-Up-Tables (LUTs), registers (FF) and memory (BRAM). Arithmetic for calculations of coordinates require limited amount of DSP blocks.

ARM processor in Zynq device has been engaged to visualize the point cloud and projection histograms as well as an interface for writing and reading control registers in the logic. The software has been written in C++ on top of PetaLinux and Xillybus infrastructure [44] for data transport between logic and DDR. This allows to access ROR coordinates from the software and to configure and monitor several parameters during the runtime.

It is important to mention that raw data (captured directly from the network receivers, before it enters the decomposition channel) is buffered on a separate data path in the design. This allows to preserve original data and perform off-line high-level analysis and an additional crosscheck between software and hardware algorithms performance.

TABLE II
FPGA RESOURCE USAGE

| Resource | Utilization | Utilization % |
|---|---|---|
| LUT | 167474 | 76.61 |
| FF | 177438 | 40.59 |
| BRAM | 432 | 79.27 |
| DSP | 208 | 23.11 |

## V. LABORATORY TESTS

A radioactive marker $^{22}$Na with activity of 1 MBq was placed inside the scanner in a center location. We have evaluated the performance of the J-PET Controller and crosschecked with GATE simulations [45]. They were performed in order to acquire theoretical estimates on hit rates and data volumes. In simulated time period of 35 seconds there was a total of 3.71 MHits registered by the scanner detecting strips including decays into 2-gamma and scatterings (Fig. 10).

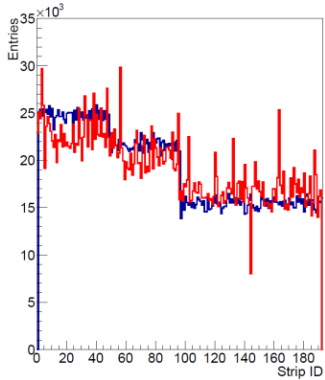

Fig. 10. Histograms of registered hits per J-PET scanner detecting strip. GATE simulations (blue) of 1 MBq inside the scanner for 35 seconds. Total number of entries is 3.71 million. Measurement performed with the J-PET scanner (red) under the same conditions. Total number of entries is 3.72 million. Non-uniform distribution on channels in measurement is due to a preliminary detector calibration.

### A. Throughput

In the measurement system, the rates of registered hits per channel varied between layers and were: 0.8 kHits/s, 0.7 kHits/s and 0.5 kHits/s for Layer 1, Layer 2 and Layer 3 respectively. Such setup produces a data stream of 85 MB per second, 450 kPackets per second that enters the J-PET Controller and delivers in total 19 MHits. 3.71 million registered hits per detector strip on lowest threshold reflects the simulated values and remains in good agreement of 0.03% difference in terms of summarized hit rates. That means no data is lost during the processing. However, the digitizing stage generates more data as a hit on a strip activates two PMTs at its ends and signal from each PMT is sampled at 4 thresholds. Therefore one hit on a strip can generate up to 8 TDC hits to process.

In order to estimate the maximum throughput of the system, a capture of processing timing has been collected (Fig. 11). It is decomposed into main tasks: gathering of input data, parsing, search for coincidence, ROR calculation and construction of output packet. One can see that the processing takes just a fraction of time between two timeslots. An output packets is constructed only in case a ROR is reconstructed, which is the case in timeslot 3 (starting at 46 µs in Fig. 11). A more detailed timing of a single timeslot processing is presented in Fig. 12 and summarized in Table III.

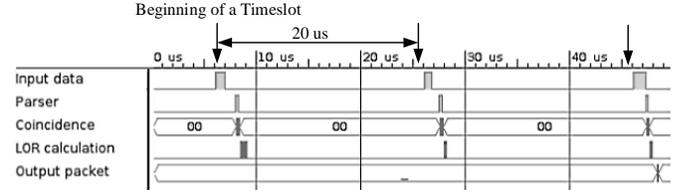

Fig. 11. Waveform captured during the measurement showing timing of the main successive processing modules. There are 3 timeslots visible within capture window, a valid ROR is reconstructed in the last one.

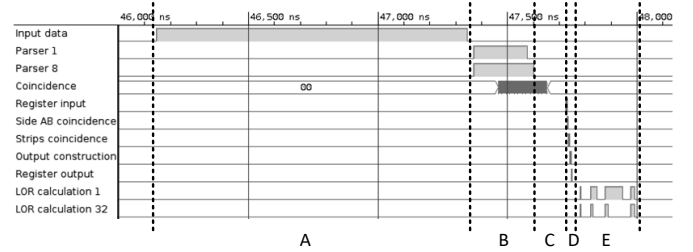

Fig. 12. Detailed timing waveform of a single timeslot processing steps. Significant part of the time is taken by the data receiver. It is due to the fact the in current implementation Gigabit Ethernet is used (125 MHz and 8 bit wide data words) and overhead in current data format. Both are subject of an upgrade.

TABLE III
PROCESSING STAGES TIMING

| Stage | Time [ns] |
|---|---|
| A. Input data reception | 1200 |
| B. Parsing | 255 |
| C. Coincidence arrays construction | 125 |
| D. Coincidence search | 25 |
| E. RORs reconstruction | 240 |
| Summarized | 1845 |
| Summarized excl. transmission | 645 |

Entire processing flow of a single timeslot containing LOR has been decomposed into 5 main stages (A-E). Total time, since reception of the first input byte until last byte is processed takes 1.845 µs. All processing stages are pipelined, therefore new input data can be received, while the previous timeslot is still being analyzed by successive stages. The Table III, helps identifying the bottleneck of the flow, which is the data reception 1.2 µs.

Maximum number of hits per second can be calculated. The timing was captured for conditions described above. 19 MHits were processed in 35 seconds, what makes 0,54 MHits per second. The longest processing stage (A) takes 1.2 µs. Time between two consecutive timeslots is 20 µs, what means that the system is able to process 16 times more data, that is 8.64 MHits per second. This value is significantly biased by the capabilities of the networking infrastructure, which can be easily upgraded. Considering only the algorithm processing time (stages B-E), the system requires 645 ns in total to process data gathered in buffers into ROR coordinates. The longest stage is data parsing (hit data extraction) which takes 255 ns. Assuming a constant stream of data, the reconstruction module could process 78 times more hits (20 µs / 0.255 µs = 78), what gives a number of 42 MHits per second.

Comparing our result to state-of-art hardware-based solutions like [46], capable of processing up to 111 million events per second, the value is more than factor 2 smaller. That can be compensated by doubling the frequency of the main clock from 200 MHz to 400 MHz (Z7045-3 limit is about 600 MHz) and assuring no timing violations will occur on combinatorial logic. Further parallelization (higher timeslot fragmentation into timebins) in expense of logic resources can also increase the throughput.

Another comparison can be made to a full-software solution, described in [4], with throughput up to 500 million events per second. That is factor 10 more than the designed system but requires four Intel Xeon X7560 CPUs, 8 cores each and 512 GB of DDR3 memory, while our system uses only one Xilinx Zynq device and 4 GB of DDR3 memory, all in a compact package. The comparison of the systems mentioned above is summarized in Table IV.

The system can produce output data stream in either raw data or in a form of list-mode data containing recovered coincidences. Is such case, we achieve a significant reduction of the data volume from almost 85 MBps down to 171 kBps that is factor 500.

TABLE IV
SYSTEMS COMPARISON

| System | J-PET Controller | DAPA [4] | LabPET II [46] |
|---|---|---|---|
| Scanner channels | 384 | 3840 | 37 k |
| Data streams | 8 | 10 | 32 |
| Hit loss | < 0.1% | < 0.1% | unknown |
| MHits per second | 42 | 500 | 111 |
| Processing units | 1x Xilinx Zynq Z7045 | 4x Intel Xeon X7560 | 17x Xilinx Virtex 5 |
| Power cons. [W] | 20 | 1 k (estim.) | 350 (estim.) |

*B. Data Quality*

The number of reconstructed RORs can be confronted with the simulation results in order to verify that no valuable data is lost in the processing pipeline. From simulation (Fig. 13), we can see that 12% (0.452M out of total 3.71M) of hits originate from 2-gamma decays that found their way to the scanner detectors. Similar value is obtained from the measurement, where 13% of all processed hits were qualified for ROR reconstruction (0.506M out of total 3.72M). There is 11% difference between simulated events qualified as LORs and extracted from the measurement. The processing logic has found more LORs due to differences in scatterings filtration and a wider time window for coincidence classification.

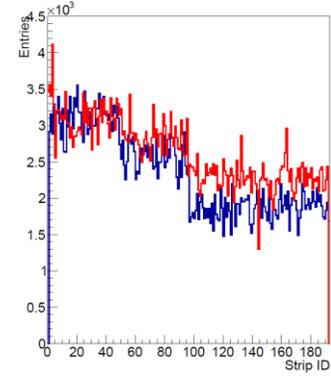

Fig. 13. Hit rates on scanner channels registered from LORs. Simulation results with registered 2-gamma decays (blue) and multiplicities on channels from reconstructed RORs by the processing pipeline (red). Due to differences in LOR classification the difference in total number of registered hits belonging to LORs is now larger at the level of 11% (452k entries in A and 506k entries in B).

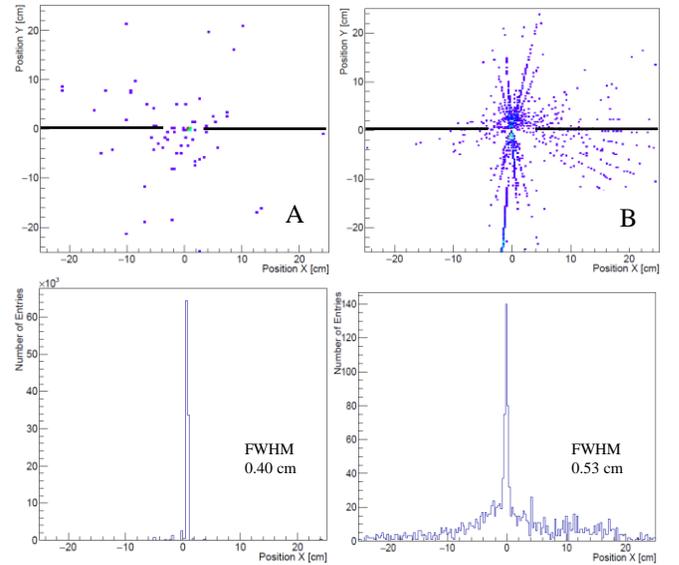

Fig. 14. Column A: Raw data analyzed with off-line, software MLEM implementation and a line profile taken at the center position, where the source was located. Column B: A point cloud of entries in 3D space calculated in the programmable logic has been projected over XY axis and a line profile (black lines) was taken at the center position (Y in range between -1 and 1 cm).

During the same measurement, 2D and 3D histograms were built out of reconstructed RORs in real-time as well as the raw data was sent to the storage for off-line processing. Simple visualization techniques were used for verification of the coordinates calculations.

The raw data, that is original 8 data streams forwarded by the J-PET Controller has been stored, reconstructed with 20 MLEM iterations [47] and presented in Fig. 14 (column A). The line profile is taken at the center position, where the 1 MBq source was located. The well pronounced main peak against almost no background is located at position 0.7 cm with an FWHM at the level of 0.4 cm.

Exactly same input data was analyzed by ROR reconstruction processing pipelines in the programmable logic and 3D point cloud has been constructed. An XY projection (Fig. 14, B) has been formed and a similar line profile applied. The main peak is located at 0.6 cm with 0.53 cm FWHM and the background level is much higher. Although no filtering is applied and just coordinates of the reconstructed points in 3D space are filled to the histograms, the results show that the calculated source coordinates match the reconstructed image.

A point cloud representation of the collected and analyzed using 3D and TOF data has been constructed and presented in Fig. 15. It shows the point source reconstructed in the center of the J-PET barrel scanner under various angles. Such data set is a validation of ROR processing algorithm and is an entry point to more sophisticated reconstruction algorithms that can be implemented in the processing device.

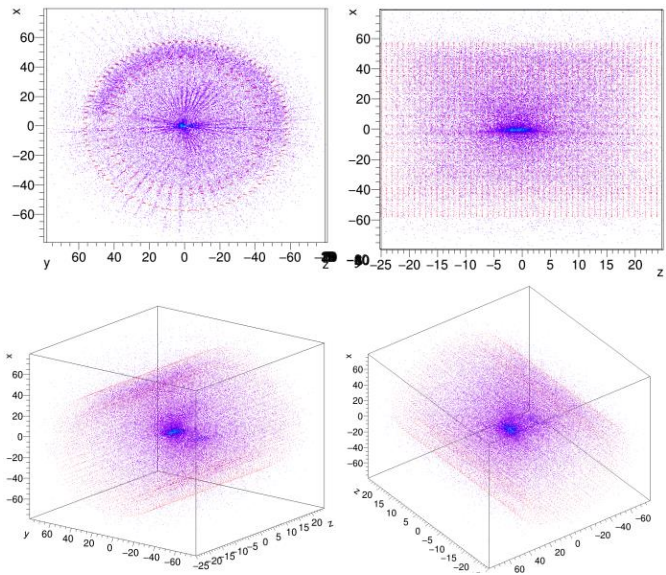

Fig. 15. 3D point cloud of the reconstructed points from the 35 second measurement, calculated in programmable logic ROR coordinates and visualized by the integrated processor. Presented are 4 different angles (top left: XY plane, top right: XZ plane, bottom left: entire scanner volume from the right corner, bottom right: entire scanner volume from top left corner). The scanner detectors locations (red points) are manually applied in order to visualize the volume.

## VI. SUMMARY AND FUTURE DEVELOPMENTS

Implemented tomographic data processing system is a first step towards fully integrated solutions for a single-chip visualization in real-time.

The processing system that is a hardware platform, firmware and software components have been developed and evaluated with the J-PET scanner. The LOR reconstruction and ROR calculation implemented in programmable logic has been positively confronted with the GATE simulation results. Each reconstructed ROR has been immediately added to a point cloud, visualized by the integrated processor. The result of such visualization has been compared to a MLEM image reconstruction, performed on the same data set.

The measurements of the described system show maximum throughput at the level of 42 MHits per second. When list-mode data output is enabled, the platform converts the raw data stream to a list of coincidences reducing the data stream by factor 500. This is especially important when the granularity of the detectors and the FOV of scanners increases producing tremendous amount of data to process.

Such systems will not overpass at the moment the quality of the off-line, iterative reconstruction algorithms but can support the original data flow significantly reducing the data volume and producing preliminary visualization. Example applications that can profit from such platforms are Image Guided Surgery systems and dosimetry in proton beam treatment, where beam monitoring is crucial.

The solution described above was implemented on a custom platform with Xilinx Zynq Z7045 and consumes about 76% resources of the device. Dynamic development of the FPGA technology brings new devices that can be applied for this purpose. Our interest is directed into migrating the solution to Xilinx ZCU102 evaluation board, which is powered by an Ultrascale+ MPSoC device. It has twice as much programmable logic resources as there are on current J-PET Controller and a much more advanced processing system facility with 4 cores ARM A53, two cores for real-time applications ARM R5 and a Mali-400 GPU. Such powerful platform is capable of computing event-by-event algorithms on the data streams provided by ROR reconstruction instantiated in programmable logic.

Increasing number of detector channels in modern scanners directs into exploration of alternative image reconstruction methods such as Origin Ensemble algorithms and its event-by-event variations. Image representations in a form of cloud of points and tetrahedral meshes are a solution for large volumes requiring voxelization in wide FOV measurements. Those two software packages will be developed after platform migration and will enable the entire solution to be widely verified according to standardized NEMA procedures [48].

The proposed solution for tomographic data processing in real-time is a proof of concept and explores the subject showing high potential for integrating all processing steps, from the raw data up to a visualization in a single-chip.

ACKNOWLEDGMENT

We acknowledge support by the National Science Centre through the grants Nos. 2016/21/B/ST2/01222, 2017/25/N/NZ1/00861 by the National Centre for Research and Development through grants Nos. INNOTECH-K1/IN1/64/159174/NCBR/12, LIDER/274/L-6/14/NCBR/2015 and by the Ministry for Science and Higher Education through grants Nos. 6673/IA/SP/2016 - IA/SP/01555/2016, 7150/E-338/SPUB/2017/I and The Foundation for Polish Science (MPD) through grant No. TEAM/2017-4/39.

B. C. Hiesmayr acknowledges gratefully the Austrian Science Fund FWF-P26783.

TRB platform together with accompanying firmware and software are developed by the TRB Collaboration (www.trb.gsi.de).

The project could be realized thanks to the support from Xilinx University Program and their donations.

The project could be realized thanks to the support from Altera University Program and their donations.

# REFERENCES

[1] A. Alessio, P. Kinahan, "PET image reconstruction", *Nucl. Med.*, vol. 1, pp. 1-22. 2006.
[2] L. A. Shepp, Y. Vardi, "Maximum Likelihood Reconstruction for Emission Tomography", *Trans. Med Imaging*, vol. 1.2, pp. 113-122, Oct. 1982
[3] H. M. Hudson, R. S. Larkin, "Accelerated image reconstruction using ordered subsets of projection data", *IEEE Trans. Med. Imaging*, vol. 13, pp. 601-609, 1994
[4] B. Goldschmidt, *et al.*, "Software-Based Real-Time Acquisition and Processing of PET Detector Raw Data", *IEEE Trans Biomedical Eng.*, vol. 63, pp. 316- 326, Feb. 2016
[5] P. Slomka, T. Pan, G. Germano, „Recent Advances and Future Progress in PET Instrumentation", *Semin. Nucl. Med.*, vol. 46, 2016
[6] S. R. Cherry, *et al.*, "Total-Body PET: Maximizing Sensitivity to Create New Opportunities for clinical Research and Patient Care", *J. Nucl. Med.*, vol. 59, pp. 3-12, Jan, 2018
[7] B. Dammak, *et al.*, "Hardware Resource Utilization Optimization in FPGA Heterogeneous MPSoC Architectures", *Microprocessors and Microsystems*, vol. 39, pp. 108-1118, June 2015
[8] J. Zhou, J. Qi, "Fast and Efficient Fully 3D PET Image Reconstruction Using Sparse System Matrix Factorization With GPU Acceleration", *Phys. Med. Biol.*, vol. 56, pp. 6739-6757, Oct. 2011
[9] A. Sitek, R. H. Huesman, G. T. Gullberg, "Tomographic Reconstruction Using an Adaptive Tetrahedral Mesh Defined by a Point Cloud", *IEEE Trans. Med. Imaging*, vol. 25, pp. 1172-1179, Sept. 2006
[10] U. Locans, *et al.*, "Real-Time Computation of Parameter Fitting and Image Reconstruction Using Graphical Processing Units", *Computer Physics Communications*, vol. 215, pp. 71-80, June 2017
[11] J. Wanza Weloli, *et al.*, "Efficiency Modeling and Exploration of 64-bit ARM Comput Nodes for Exascale", *Microprocessors and Microsystems*, vol. 53, pp. 68-80, Aug. 2017
[12] D. Sun, *et al.*, "Radioimmunoguided Surgery (RIGS), PET/CT Image-Guided Surgery and Fluorescence Image-Guided Surgery: Past, Present and Future", *J. Surgical Oncology*, vol. 96, pp. 297-308, 2007
[13] S. Binet, *et al.*, "Construction and First Tests of an in-beam PET Demonstrator Dedicated to the Ballistic Control of Hadrontherapy Treatments With 65 MeV Protons", *IEEE Trans. Radiation and Plasma Medical Sciences*, vol. 2, pp. 51-60, Jan. 2018
[14] H. Fayad, F. Lamare, T. Merlin, D. Visvikis, "Motion Correction using anatomical information in PET/CT and PET/MRI Hybrid Imaging", *Q. J. Nucl. Med. Mol. Imaging*, vol. 60, pp. 12-24, Mar. 2016
[15] Liu X., Liu S., An Q., "A time-over-threshold technique for PMT signals processing", *Nucl. Sci. and Technique*s, vol. 18, pp. 164-171, June 2007
[16] G. B. Ko, *et al.*, "Development of FPGA-based Coincidence Units with Veto Function", *Biomed. Eng. Lett.*, vol. 1, pp. 27-31, Dec. 2010
[17] A. P. Dempster, N. M. Laird, D. B. Rubin, "Maximum Likelihood from Incomplete Data via the EM Algorithm", *J. of the Royal Statistical Society series B.*, vol. 39, pp. 1-39, 1977
[18] C. Wulker, A. Sitek, S. Prevrhal, "Time-of-flight PET Image Reconstruction Using Origin Ensembles", *Phys. Med. Biol.*, vol. 60, pp. 1919-44, Mar. 2015
[19] S. S. Huh, L. Han, W. L. Rogers, N. J. Clinthorne, "Real time image reconstruction using GPUs for a surgical PET imaging probe system", *IEEE Nuclear Science Symposium Conference Record (NSS/MIC)*, 2009
[20] L. Shi, W. Liu, H. Zhang, Y. Xie, D. Wang, "A survey of GPU_based medical image computing techniques", *Quantative Imaging in Medicine and Survey*, vol. 2(3), pp. 188-206, 2012
[21] G. Saldaña-González, *et al.,* "2D image reconstruction with a FPGA-based architecture in a gamma camera application.", *Electronics, Communications and Computer (CONIELECOMP), 2010 20th International Conference on. IEEE*, 2010.
[22] N. Gac, S. Mancini, M. Desvignes, D. Houzet, "High speed 3D Tomography on CPU, GPU and FPGA", *EURASIP Journal on Embedded Systems*, 930250, 2008
[23] M. Haselman, *et al.*, "FPGA-based Front-End Electronics for Positron Emission Tomography", *FPGA*, vol. 7, pp. 93-102, Feb. 2009
[24] P. Moskal, *et al.*, "Novel detector systems for the Positron Emission Tomography", *Bio-Algorithms and Med-Systems*, vol. 7, pp. 73-78, 2011
[25] P. Moskal, *et al.*, "Time resolution of the plastic scintillator strips with matrix photomultiplier readout for J-PET tomograph", *Phys. Med. Biol.*, vol. 61, pp. 2025-2047, 2016
[26] P. Moskal, *et al.*, "A novel method for the line-of-response and time-of-flight reconstruction in TOF-PET detectors based on library of synchronized model signals", *Nucl. Inst. And Meth. A*, vol. 775, pp. 54-62, 2015
[27] P. Moskal, *et al.*, "Test of a single module of the J-PET scanner based on plastic scintillators", *Nucl. Instr. And Meth. A*, vol. 764, pp. 317-321, 2014
[28] L. Raczyński, et al., "Calculation of time resolution of the J-PET tomograph using the kernel density estimation", *Phys. Med. Biol.*, vol. 62, 2017
[29] Sz. Niedzwiecki, *et al.*, "J-PET: A New Technology for the Whole-Body PET Imaging", *Acta Phys. Polon. B.*, vol. 48, pp. 1567-1576, 2017
[30] A. Sitek, "Representation of photon limited data in emission tomography using origin ensembles", *Phys. Med. Biol.*, vol. 53, pp. 3201-3216, June 2008
[31] G. Korcyl, *et al.*, "Sampling FEE and Trigger-less DAQ for the J-PET Scanner", *Acta Phys. Pol. B*, vol. 47, pp. 491-496, 2016
[32] G. Korcyl, P. Moskal, M. Kajetanowicz, M. Pałka, „A system for acquisition of tomographic measurement data", WO/2015/028594, Publ. 05.0
[33] G. Korcyl, "A novel data acquisition system based on fast optical links and universal readout boards", Ph.D. dissertation, Fac. Comp. Science, Electr. And Telecom., Univ. of Science and Tech., Kraków, Poland, 2015.
[34] M. Traxler *et al.*, "A compact system for high precision time measurements (< 14 ps RMS) and integrated data acquisition for a large number of channels", *JINST* vol. 6 C12004, Dec. 2011.
[35] C. Ugur, G. Korcyl, J. Michel, M. Penschuk, M. Traxler, "264 Channel TDC platform applying 65 channel high precision (7.2 ps RMS) FPGA based TDCs", *IEEE Nordic-Mediterranean Workshop on Time-to-Digital Converters (NoMe TDC)*, pp. 1-5 Oct. 2013.
[36] M. Pałka, et al., "Multichannel FPGA based MVT system for high precision time (20 ps RMS) and charge measurement", JINST 12 P08001, 2017
[37] L. Raczyński, *et al.*, "Application of the compress sensing theory for improvement of the TOF resolution in a novel JPET instrument", *Nukleonika*, vol. 61, pp 35-39, 2016
[38] L. Raczyński, *et al.*, "Compressive sensing of signals generated in plastic scintillators in a novel J-PET instrument", *Nucl. Instr. And Meth. A*, vol. 786, pp. 102-112, 2015
[39] L. Raczyński, *et al.*, "Novel method for hit-position reconstruction using voltage signals in plastic scintillators and its application to the Positron Emission Tomography", *Nucl. Instr. And Meth. A*, vol. 764, pp. 186-192m 2014
[40] B. Goldschmidt, *et al.*, "Towards Software-Based Real-Time Singles and Coincidence Processing of Digital PET Detector Raw Data", *IEEE Trans. Nucl. Sci.*, vol. 60, pp. 1550-1559, June 2013
[41] P. Strzempek, "Development and evaluation of a signal analysis and a readout system of straw tube detectors for the PANDA spectrometer", Ph.D. dissertation, Fac. Phys. Astr. And App. Comp. Science, Jagiellonian Univ., Kraków, Poland, 2017.
[42] C. Ugur, E. Bayer, N. Kurz, M. Traxler, "Implementation of a High Resolution Time-to-Digital Converter in a Field Programmable Gate Array", in *Proc. 50th International Winter Meeting on Nuclear Physics*, Bormio, Italy 2012.
[43] "Vivado Design Suite User Guide – High-Level Synthesis", Xilinx, UG902, 2014,
[44] "Xillybus – product brief", Xillybus.com, 2018
[45] S. Jan, *et al.,* "GATE: a simulation toolkit for PET and SPECT", *Phys. Med. Biol.,* vol. 49, pp. 4543-4561, 2004
[46] L. Njejimana, *et al.*, "Design of a real-time FPGA-based DAQ architecture for the LabPET II, an APD-based scanner dedicated to small animal PET imaging", *2012 18th IEEE-NPSS Real Time Conf.*, pp. 1-5, 2012
[47] A. Strzelecki, "Image reconstruction and simulation of strip Positron Emission Tomography scanner using computational accelerators", Ph.D. dissertation, Inst. of Fundamental Technological Research PAN, Warsaw, Poland, 2016.
[48] National Electrical Manufacturers Association, "NEMA Standards Publication NU 2-2007: Performance Measurements of Positron Emission Tomographs", *NEMA*, 2007